\documentclass[letterpaper]{article} %
\usepackage[preprint]{aaai2027}  %
\usepackage[hyphens]{url}  %
\usepackage{graphicx} %
\usepackage{natbib}  %
\usepackage{caption} %
\usepackage{algorithm}
\usepackage{algorithmic}

\usepackage{newfloat}
\usepackage{listings}
\DeclareCaptionStyle{ruled}{labelfont=normalfont,labelsep=colon,strut=off} %
\floatstyle{ruled}
\newfloat{listing}{tb}{lst}{}
\floatname{listing}{Listing}

\usepackage{booktabs}
\usepackage{array}
\usepackage{amsmath}
\usepackage{amssymb}

\newcommand{\atom}{\mathcal{A}}
\newcommand{\template}{\mathcal{T}}

\title{RangeFactory: Scalable Construction of Multi-Hop Cyber Ranges}
\author{
    Hanlin Jiang\textsuperscript{\rm 1},
    Puyi Wang\textsuperscript{\rm 1},
    Jiandong Jin\textsuperscript{\rm 1},
    Shaofei Li\textsuperscript{\rm 1},
    Zhan Shen\textsuperscript{\rm 1},
    Pengli Wang\textsuperscript{\rm 1},
    Ziming Wang\textsuperscript{\rm 1},
    Yifeng Cai\textsuperscript{\rm 1},
    Ning Jia\textsuperscript{\rm 2},
    Yuxin Ren\textsuperscript{\rm 2},
    Peng Jiang\textsuperscript{\rm 3},
    Yao Guo\textsuperscript{\rm 1},
    Ding Li\textsuperscript{\rm 1}\corresponding
}
\affiliations{
    \textsuperscript{\rm 1}Key Laboratory of High Confidence Software Technologies, Peking University, Ministry of Education\\
    \textsuperscript{\rm 2}Huawei Technologies Co., Ltd.\\
    \textsuperscript{\rm 3}Southeast University
}

\begin{document}

\maketitle

\begin{abstract}
Real-world cyberattacks often require sustained progress across multiple hosts
and network segments, making multi-hop cyber ranges essential infrastructure
for studying and improving LLM agents' ability to sustain complete attack chains.
Prior work has scaled isolated vulnerability tasks and constructed
multi-host scenarios from manually specified vulnerability semantics. However, they are still unable to automatically orchestrate the growing supply of vulnerability
environments into end-to-end validated multi-hop ranges. To this end, we present
\textbf{RangeFactory}, an automated cyber-range orchestration framework that
constructs multi-hop cyber ranges at scale from isolated vulnerability
environments. RangeFactory formulates range construction as dependency
resolution: it extracts dependency information from agents' actual attacks
against real vulnerabilities, resolves known dependencies through
template-guided orchestration, and uses end-to-end attack execution to validate
runtime dependencies that emerge after composition.

Using RangeFactory, we construct \textbf{RangeBench} with 1,148 validated range
instances spanning 287 distinct attack chains and evaluate frontier attack
agents across attack depth, network scale, and task information. Among runs
that compromise the entry vulnerability, 24.5--47.0\% still fail to complete
the remaining attack path, revealing a substantial sustained-compromise gap
between establishing an initial foothold and completing a multi-hop attack.
RangeFactory further produces a corpus of 5,541 outcome-annotated multi-hop
attack trajectories, providing execution data for attack-process analysis and
future agent training.
\end{abstract}

\section{Introduction}

Large language model agents are evolving from executing isolated instructions
to autonomously completing long-horizon tasks in complex environments. This
trend is extending agent autonomy to cybersecurity, where real-world
intrusions often require sustained progress across multiple network
positions~\cite{PentestGPT,agentsautonomouslyexploit}. A cyber range is an
interactive network environment for executing and validating cyberattacks; a
multi-hop cyber range further requires an attacker to use footholds established
by earlier compromises to reach targets in distinct network segments. Such
environments therefore provide the infrastructure for systematically studying
whether attack agents can sustain multi-hop compromise~\cite{folkerts}.

Prior work has developed two classes of resources for evaluating attack
agents. One line scales the conversion of real vulnerabilities into executable
single-vulnerability tasks, but each task ends at an isolated exploit and
cannot evaluate sustained compromise across network positions~\citep{CVEFactory}.
Another line uses multi-host cyber ranges to evaluate end-to-end attacks, but
existing benchmarks depend on a small collection of expert-designed, fixed
scenarios~\citep{AgentCyberRange}. Automated scenario-generation systems can
expand environment deployment, yet still require experts to specify
cross-vulnerability dependencies in advance~\citep{AutoCyberExerciseScenarios}.
As a result, existing works cannot automatically set up the environments that support complete attack chains.
Therefore, this paper addresses how to
automatically transform the growing supply of single-vulnerability
environments into multi-hop cyber ranges with executable end-to-end attack
paths.

The central challenge in this transformation is automatically making independently
exploitable vulnerabilities execute consecutively within one attack. This
requires resolving both capability and runtime dependencies. A capability
dependency determines whether adjacent attacks can connect; for example,
file-read access cannot support a subsequent exploit that requires command
execution. A runtime dependency determines whether the concrete attack
procedure can operate from its new foothold; for example, a network policy may
block an exploit's callback~\citep{NGUYEN2025104140}. Directly composing
vulnerability environments and executing every combination creates a rapidly
growing candidate space. Dependency-guided orchestration can eliminate known
dependency conflicts before deployment, while conditions introduced by joint
deployment still require complete attack execution. Scalable construction
therefore first composes candidates whose dependencies can connect and then
executes each end to end to confirm that the path remains feasible.

Based on this observation, we present \textbf{RangeFactory}, a multi-agent
system that automatically transforms isolated vulnerability environments into validated
multi-hop cyber ranges. RangeFactory views range construction as a
dependency-resolution process: it characterizes dependencies through actual
attacks, resolves them through scenario orchestration, and validates the
resulting runtime conditions through end-to-end execution. Atomization agents
attack each CVE environment and probe its post-exploitation capabilities,
producing a CVE Capability Atom with dependency information and a reusable
attack procedure. Orchestration agents generate attack-path templates and bind
compatible Atoms, after which Validation agents execute the complete attack
against hidden objectives. Diagnostic feedback supports attack retries and
subsequent path construction. Every retained range thereby contains a path
successfully exercised during construction.

Using 239 CVE Capability Atoms and 20 enterprise-style network configurations,
RangeFactory produces 1,840 candidate range instances. End-to-end validation
retains 1,148 instances spanning 287 distinct ordered attack chains, for a
validation rate of 62.4\%. We organize them into \textbf{RangeBench} and
evaluate frontier attack agents across attack depth, network scale, and task
information. The results reveal a substantial sustained-compromise gap:
success on an entry vulnerability does not reliably translate into completing
the attack when the same path is extended across additional network layers.
Beyond evaluation, RangeFactory constructs a corpus of 5,541
outcome-annotated multi-hop interaction trajectories from the same ranges.
Validated demonstrations and autonomous outcomes share verifier-derived
milestone annotations that may support future attack-agent training.

Our contributions are summarized as follows:
\begin{itemize}
    \item \textbf{RangeFactory.} We introduce an automated multi-agent
    framework that derives vulnerability dependencies through actual attacks
    and combines orchestration with end-to-end validation to construct
    multi-hop cyber ranges at scale.
    \item \textbf{RangeBench and Empirical Findings.} We construct RangeBench
    with 1,148 validated range instances spanning 287 distinct attack chains
    and systematically study how attack depth, network scale, and task
    information affect frontier attack agents, revealing a substantial gap
    between early compromise and sustained multi-hop compromise.
    \item \textbf{Outcome-Annotated Attack Trajectories.} We collect a corpus
    of 5,541 multi-hop trajectories pairing validated demonstrations with
    autonomous outcomes on the same ranges, with verifier-derived milestone
    annotations that may support future attack-agent training.
\end{itemize}

\section{Related Work}

\paragraph{Executable Cybersecurity Tasks and Agent Data.}
Recent work has transformed real vulnerabilities into executable tasks,
grounding cyber-agent evaluation in actual interaction~\citep{CVEBench,CyberGym}.
Automated vulnerability reproduction further converts sparse CVE metadata into
complete single-vulnerability task packages~\citep{CVEGenie,CVEFactory}. Complementary benchmark efforts package
professional CTF and penetration-testing challenges as reproducible
interactive environments~\citep{NYUCTFBench,Cybench}. Beyond
benchmark construction, cybersecurity-agent data have been scaled from CTF artifacts: runtime-free
methods synthesize trajectories from writeups, while execution-grounded
platforms collect verified rollouts from containerized
challenges~\citep{CyberZero,CTFDojo}. These resources remain centered on
individual vulnerabilities or challenges. Multi-host cyber-range benchmarks
evaluate how agents expand an initial foothold across a network, but rely on a
small, fixed collection of ranges~\citep{AgentCyberRange,liu2025pacebench}. RangeFactory
addresses this environmental bottleneck and collects outcome-annotated
multi-hop interactions as a downstream artifact of range construction and
evaluation.

\paragraph{Automated Cyber-Range Construction.}
A separate line of work automates cyber-range construction. Attack-graph
methods derive multi-host attack chains from formal system
models~\citep{MulVAL}. Cyber-range generators instantiate expert-authored
scenario models~\citep{CyRIS,SecGen}, while model-driven approaches support
automated deployment validation~\citep{VSDL,CRACK}. LLM-based systems further translate natural-language range requirements into
platform-specific configurations and deployment
artifacts~\citep{ARCeR,AICyberRangeAssistant}, while constraint-based synthesis
uses model finding and capability semantics to generate enterprise scenarios
at scale~\citep{AutoCyberExerciseScenarios}.
These approaches automate cyber-range construction once authored scenario descriptions or predefined composition semantics are available, but adding
a new vulnerability still requires its attack dependencies to be specified in
advance. RangeFactory obtains these dependencies from successful attacks
against existing vulnerability environments, uses them to orchestrate
candidate chains, and then executes each deployed chain end to end to detect unresolved runtime incompatibilities introduced by composition.

\section{System Design}

\subsection{Insight and Overview}

Deploying independently exploitable CVE environments together does not yield a valid multi-hop attack chain unless both capability and runtime dependencies hold. A capability dependency concerns what an attack requires and produces: a preceding compromise must provide the capability needed by the next exploit. File-read access, for example, cannot support an exploit that requires command execution. A runtime dependency concerns whether the concrete attack procedure can operate from its new execution position. A blocked reverse connection or a missing tool on the compromised host can prevent exploitation even when the capability interface matches. Either violation breaks the consecutive execution of otherwise exploitable CVEs.

RangeFactory views multi-hop range construction as dependency resolution: actual attacks characterize dependencies, orchestration resolves them, and end-to-end execution confirms candidate-chain executability. Single-vulnerability attacks reveal the capabilities required and produced and record the runtime requirements observed during exploitation. Orchestration connects consecutive capabilities and uses the network template to constrain known runtime dependencies. Because the new footholds and joint deployment can introduce conditions that were absent from the single-vulnerability environment, RangeFactory then executes the complete attack to expose remaining conflicts. This division prunes clearly infeasible compositions before deployment while retaining only chains successfully exercised after composition.

\begin{figure*}[t]
    \centering
    \includegraphics[width=\textwidth]{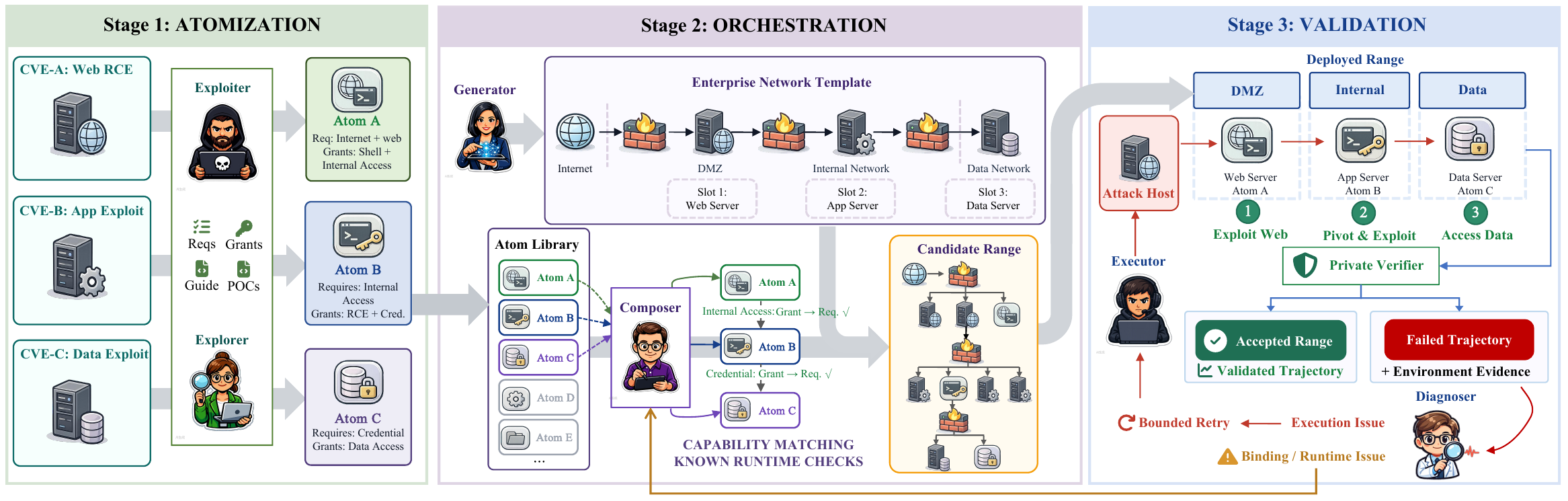}
    \caption{Overview of RangeFactory's multi-agent workflow for constructing
    validated multi-hop cyber ranges.}
    \label{fig:overview}
\end{figure*}

Figure~\ref{fig:overview} presents RangeFactory's multi-agent workflow.
Atomization produces CVE Capability Atoms, Orchestration binds them into
candidate ranges, and Validation exercises the complete attack before
acceptance. Diagnostic feedback is routed to attack execution for bounded
retry or to chain construction as persistent incompatibility evidence.

\subsection{CVE Capability Atom}

RangeFactory represents each successfully attacked vulnerability environment as a CVE Capability Atom, which binds the deployable environment to dependency information and execution artifacts. Table~\ref{tab:atom-contract} summarizes its four fields. \texttt{exploit\_access} records the structured conditions under which the exploit was exercised. \texttt{capability\_grants} binds each verified post-compromise capability to the compromised principal and its evidence; only verified grants participate in orchestration. A grant's host scope is assigned when the Atom is bound to a template position. \texttt{exploit\_guide} preserves the validated procedure and observed runtime requirements, while referenced PoC files are packaged as \texttt{poc\_materials}. The interface supports dependency matching, and the execution artifacts allow the vulnerability to be exercised again in a composed range.

\begin{table}[t]
\centering
\footnotesize
\setlength{\tabcolsep}{3pt}
\renewcommand{\arraystretch}{1.05}

\begin{tabular}{@{}
    >{\raggedright\arraybackslash}p{0.32\columnwidth}
    >{\raggedright\arraybackslash}p{0.62\columnwidth}
    @{}}
\toprule
\multicolumn{1}{c}{\textbf{Field}} &
\multicolumn{1}{c}{\textbf{Semantics}} \\
\midrule

\multicolumn{2}{@{}l}{\textit{Capability Interface}} \\[-1pt]

\texttt{exploit\_access}
&
Structured preconditions: attack vector, privilege, and service protocol/port.
\\[2pt]

\texttt{capability\_}\newline
\texttt{grants}
&
Verified post capability from a fixed ontology, with principal and evidence.
\\

\midrule

\multicolumn{2}{@{}l}{\textit{Execution Artifacts}} \\[-1pt]

\texttt{exploit\_guide}
&
Successful procedure with tool, material, authentication, and callback requirements.
\\[2pt]

\texttt{poc\_materials}
&
Exploit artifacts referenced by \texttt{exploit\_guide}.
\\

\bottomrule
\end{tabular}
\normalsize
\caption{Description of CVE Capability Atom fields.}
\label{tab:atom-contract}
\end{table}

The capability ontology represents command execution
\path{execute_command}, file access \path{read_file} and
\path{write_file}, network position \path{network_vantage},
credential discovery \path{read_credential}, and authentication
\path{authenticate}. For example, when
\path{exploit_access} requires an internal service, the preceding
compromise must provide \path{network_vantage}. A TCP callback requirement
is instead recorded in \path{exploit_guide} for runtime compatibility
checking. Capability matching uses only grants confirmed by probes, avoiding
chains supported by an Agent's unverified inference. Runtime dependencies
vary with the composed environment and cannot be completely exposed by a
single-vulnerability attack, so the Atom records observed requirements and
leaves the remainder to end-to-end validation.

\subsection{Atomization}

\paragraph{Exploiter.}
RangeFactory deploys a reproducible vulnerability environment, configures a
private objective aligned with the vulnerability impact, and provides the
Exploiter with its reproduction context and available exploit materials. The
Exploiter adapts the attack according to environment feedback, while an
independent verifier confirms success against the private objective. After a
successful attack, it distills the procedure, exploit conditions, and observed
runtime dependencies into \texttt{exploit\_guide}. This establishes for every
Atom an execution-confirmed attack for subsequent range construction.

\paragraph{Explorer.}
Reaching the vulnerability objective confirms its expected security impact;
orchestration also needs to know which subsequent actions that impact enables.
The Explorer therefore reuses the Exploiter's environment and established
foothold, selecting capability probes according to the available access
mechanism and probe feedback. For example, it directs the compromised host to
a designated endpoint to confirm \texttt{network\_vantage}. Only capabilities
confirmed by probes and their corresponding principals enter
\texttt{capability\_grants}. This process converts heterogeneous vulnerability
impacts into a uniform interface for subsequent orchestration.

\subsection{Orchestrator}

\paragraph{Generator.}
The Generator transforms public architecture references into parameterized
enterprise-style templates. Each template defines the network structure and
communication policies and marks an attack-chain skeleton from an external
entry point to the final objective. It reserves one CVE slot in every
traversed layer and independently parameterizes background business assets to
control the search space. For example, a three-tier template can traverse a
DMZ, an internal application network, and a data network. The template fixes
chain structure and network-dependent constraints while keeping Atom selection
open.

\paragraph{Composer.}
The Composer takes a scenario template and a library of verified Atoms as
input and constructs candidate bindings in attack-chain order. For each CVE
slot, it retrieves Atoms whose service requirements are compatible with the
template position and checks whether the capabilities accumulated at preceding
positions satisfy the next Atom's \texttt{exploit\_access}. After each binding,
the Composer associates the Atom's \texttt{capability\_grants} with the current
network position and updates the capability set through fixed closure rules.
For example, verified command execution can support file access and network
vantage under the same principal. A partial chain is pruned as soon as a
dependency is violated, after which the Composer backtracks to another Atom.

Beyond capability dependencies, the Composer combines the network template
with \texttt{exploit\_guide} to check runtime requirements that can be
determined before deployment, such as the attack's execution position and
material delivery. The concrete exploit procedure and runtime dependencies
that emerge only in the composed environment are left to end-to-end execution.
Once all slots are bound, the Composer forms an attack graph whose nodes are
Atom-bound template positions and whose edges are dependency-satisfying attack
transitions. When the candidate set is large, it prioritizes chains that expand
coverage of Atom and asset configurations. The output records Atom bindings,
capability transitions, and deployment parameters for range instantiation and
validation.

\subsection{Validation}

\paragraph{Executor.}
After deployment, RangeFactory checks that all nodes and services operate
under the network configuration defined by the template. The Executor
accesses each Atom's execution artifacts, organizes the per-hop exploits into
an end-to-end attack, and adapts them as the foothold and runtime environment
change. A verifier independently checks the ordered private objectives hidden
from the Executor. Only a candidate that completes the end-to-end attack is
accepted; the complete trajectory and verification result are retained, while
failed executions are passed to the Diagnoser.

\paragraph{Diagnoser.}
The Diagnoser locates the failed step from the trajectory and environment
evidence and distinguishes attack execution from chain composition failures.
An execution failure is returned to the Executor for a bounded retry that
preserves the current context. An unsatisfied runtime dependency instead
filters the candidate and produces an incompatibility record associated with
the binding conditions. The Composer retrieves these records during later
orchestration, allowing evidence from validated failures to improve subsequent
candidate generation.

\section{Experiment}

\subsection{Scalable and Reliable Construction of Multi-Hop Ranges}

\paragraph{Setup.}
To evaluate whether RangeFactory can construct multi-hop cyber ranges at
scale, we automatically screen and deduplicate vulnerability resources from
Vulhub and CVE-Factory~\citep{Vulhub,CVEFactory}. Eligible resources must
provide a deployable network service and allow the vulnerability impact to be
verified through network interaction. This process yields 257 Atomization
candidates, of which 239 are successfully converted into CVE Capability Atoms
(93.0\%); the remaining failures arise from missing build artifacts or
unstable reproduction. These Atoms cover 122 products and CVEs disclosed
between 2010 and 2026. Given
public architecture references and the requested scale, the Generator
produces five three-tier enterprise-style templates without manual CVE binding
or chain construction. Each places one vulnerable node per tier, yielding an
entry compromise and two pivots. Although the representation accepts
variable-length attack chains, three tiers provide matched one-, two-, and three-hop
measurements within the same complete range. Four background-asset scales
produce 20 configurations spanning small to medium-sized networks. All
construction agents use DeepSeek-V4-Pro; validation receives the Atom
execution artifacts, 300 turns, and a one-hour timeout. Source selection and
the choice of architecture references are one-time dataset preparation.
RangeFactory instantiates private objectives and capability probes from
reusable adapters, then performs Atomization, CVE binding, deployment, and
end-to-end validation without per-range expert annotation or chain design.

\begin{table}[t]
\centering
\footnotesize
\renewcommand{\arraystretch}{1.12}
\begin{tabular*}{\columnwidth}{@{\extracolsep{\fill}}cc@{}}
\toprule
\textbf{Stage} & \textbf{Transformation} \\
\midrule
Atomization
  & 257 candidates $\rightarrow$ 239 Atoms (93.0\% successful) \\
Orchestration
  & 23,789 prop. $\rightarrow$ 1,840 cand. (92.3\% filtered) \\
Validation
  & 1,840 cand. $\rightarrow$ 1,148 ranges (62.4\% validated) \\
\bottomrule
\end{tabular*}
\normalsize
\caption{Stage-wise construction and validation results. Prop. and cand.
denote proposals and candidates.}
\label{tab:range-construction}
\end{table}

\paragraph{Construction Scale and Coverage.}
Given 239 CVE Capability Atoms and 20 scenario configurations, the Composer
examines 23,789 CVE binding proposals. It rejects 21,949 proposals whose
capability or runtime dependencies cannot be satisfied and produces 1,840
candidate range instances, pruning 92.3\% before deployment. The candidates form 460
matched groups that instantiate the same template and Atom chain at four
network scales. End-to-end validation retains 287 groups, each associated
with a distinct ordered Atom chain, yielding 1,148 range instances that cover
213 Atoms, 113 products, and all 20 configurations. Each chain is instantiated
in four matched network contexts whose background assets differ in scale and
composition. The resulting benchmark therefore separates attack-chain
diversity from variation in the surrounding network.

\paragraph{Execution Grounding.}
RangeFactory deploys and validates the 1,840 candidate range instances, of which 1,148
complete an end-to-end attack and pass private-objective verification, for an
overall validation success rate of 62.4\%. Each candidate directly reuses the
vulnerability environments and execution artifacts packaged in its Atoms;
instantiating it and completing the first validation pass takes 16.2 minutes
on average. Because Atomization is performed once for each vulnerability
environment, its outputs can be reused across many candidate ranges without
repeating the single-vulnerability attack.

\paragraph{Construction Quality.}
Scale alone does not establish construction quality. We therefore randomly
sample 150 accepted ranges for an independent audit that checks whether the
injected vulnerability preserves its original security impact and whether the
Atom's access and capability fields agree with the observed attack states.
Among the audited ranges, 146 (97.3\%) preserve the source vulnerability
behavior and 142 (94.7\%) agree with the Atom annotations. Separately, every
retained range has an attack witness that completes its ordered private
objectives; this constructive evidence does not assert the absence of other
valid attack chains. The audit supports preservation of vulnerability behavior and
dependency semantics.

\subsection{RangeBench: Evaluating Multi-Hop Attack Agents}

\paragraph{Benchmark Design.}
We organize 1,148 end-to-end validated range instances---287 distinct ordered
attack chains, each instantiated in four matched network contexts---into
RangeBench to study sustained execution, exploration, and prior knowledge. A private
objective on each vulnerable node defines matched one-, two-, and three-hop
success within the same rollout: reaching the first objective is one-hop
success, reaching the first two in order is two-hop success, and reaching all
three is end-to-end success. Deeper layers become reachable only after the
preceding foothold, so the measurements share an identical entry vulnerability
and chain prefix. Each chain also has none, low, medium, and high scale variants
with approximately 7, 12, 31, and 50 visible nodes. Atom bindings and task
information remain fixed within each group; only background business assets
vary. Finally, L0 provides the entry address and objective, L1 adds network
topology, and L2 adds target services, CVE mappings, and required credentials.
No level provides PoCs or Exploit Guides.

\paragraph{Evaluation Setup.}
We evaluate Kimi-K3, GLM-5.2, DeepSeek-V4-Pro, and GPT-5.6-Luna through a
common task interface. We execute one rollout for every model--range--condition
pair and expose the same task prompt and attack tools, with a budget of 300
turns and one hour per run. Kimi-K3 uses temperature 1, the minimum supported
by its endpoint; the other models use temperature 0, and all other sampling
parameters retain their provider defaults. Transient API errors are retried
and infrastructure failures rerun; refusals, early termination, and budget
exhaustion count as failures. Evaluation agents cannot access Atom execution
artifacts. The results characterize pass@1 across the complete balanced
benchmark rather than within-task sampling variance.

\paragraph{Metrics.}
End-to-end success is the fraction of runs whose final objective is verified.
Hop-$k$ success is the fraction that consecutively completes the first $k$
objectives. Chain progress divides the deepest verified hop in each run by the
required chain length and averages this fraction across runs. We report median
tool calls and wall-clock time as interaction cost: the former measures the
amount of agent exploration, while the latter captures the resulting
execution cost.

\begin{table}[t]
\centering
\footnotesize
\begin{tabular}{@{}ccccc@{}}
\toprule
\textbf{Model} &
\shortstack{\textbf{End-to-End}\\\textbf{Success}} &
\shortstack{\textbf{Chain}\\\textbf{Progress}} &
\shortstack{\textbf{Tool}\\\textbf{Calls}} &
\shortstack{\textbf{Time}\\\textbf{(min)}} \\
\midrule
Kimi-K3         & \textbf{55.7} & \textbf{64.6} & 138 & 44.6 \\
GLM-5.2         & 41.2          & 51.2          & 145 & 46.8 \\
DeepSeek-V4-Pro & 36.2          & 46.2          & 152 & 48.5 \\
GPT-5.6-Luna    & 23.5          & 32.9          & 86  & 27.4 \\
\bottomrule
\end{tabular}
\normalsize
\caption{Overall performance on RangeBench under L2 task information.
End-to-end success and chain progress are percentages; tool calls and time are
medians.}
\label{tab:rangebench-overall}
\end{table}

\begin{figure*}[t]
\centering
\includegraphics[width=\textwidth]{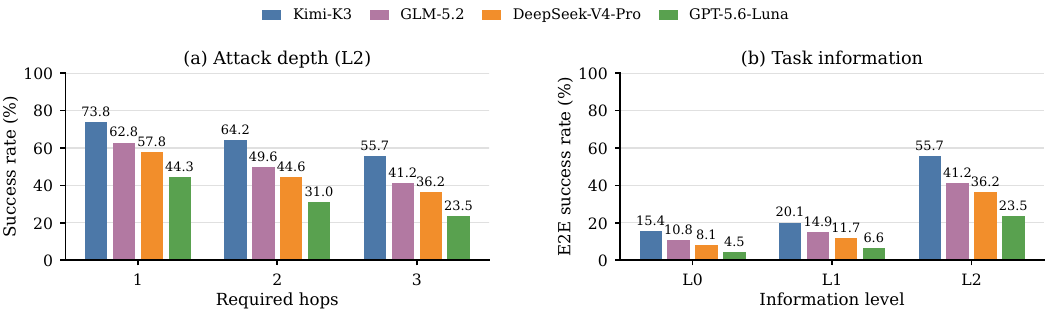}
\caption{RangeBench factor analysis across four attack agents. (a) Success at
successive attack depths under L2 information. (b) End-to-end success under
different task-information levels.}
\label{fig:rangebench-dimensions}
\end{figure*}

\paragraph{Overall Performance.}
Table~\ref{tab:rangebench-overall} shows substantial separation among the
models even under L2, where target services and their CVEs are provided.
Kimi-K3 leads at 55.7\% end-to-end success, followed by GLM-5.2 at 41.2\%
and DeepSeek-V4-Pro at 36.2\%. GPT-5.6-Luna reaches 23.5\%, with median
interaction costs of 86 tool calls and 27.4 minutes, substantially below the
other models. Its lower success therefore coincides with earlier termination
and less extensive exploration on difficult attacks.

\paragraph{Attack Depth.}
Figure~\ref{fig:rangebench-dimensions}(a) uses the private objectives placed
along the three network layers to evaluate the same executions at successive
depths. Because the measurements share an identical chain prefix and deeper
services become reachable only after the preceding foothold, differences
across depths cannot be attributed to a harder entry vulnerability. Kimi-K3
reaches the first, second, and third objectives in 73.8\%, 64.2\%, and 55.7\%
of trials, respectively. GLM-5.2 declines from 62.8\% to 41.2\%,
DeepSeek-V4-Pro from 57.8\% to 36.2\%, and GPT-5.6-Luna from 44.3\% to
23.5\%. Conditioned on reaching the first objective, only 75.5\%, 65.6\%,
62.6\%, and 53.0\% of runs complete the full chain for Kimi-K3, GLM-5.2,
DeepSeek-V4-Pro, and GPT-5.6-Luna, respectively. Thus, 24.5--47.0\% of runs
that establish an initial foothold still fail on later compromises, revealing
a sustained-compromise gap that accumulates with depth.

\paragraph{Task Information.}
Figure~\ref{fig:rangebench-dimensions}(b) compares all four models under the
three information levels. End-to-end success increases from 15.4\% under L0
and 20.1\% under L1 to 55.7\% under L2 for Kimi-K3. GLM-5.2 improves from
10.8\% and 14.9\% to 41.2\%, while DeepSeek-V4-Pro rises from 8.1\% and
11.7\% to 36.2\%. GPT-5.6-Luna follows the same pattern at 4.5\%, 6.6\%,
and 23.5\%. L0 and L1 produce only a modest difference, whereas L2
substantially improves all four models. The best model nevertheless fails on
nearly half of the ranges, showing that richer task information does not
eliminate the difficulty of sustained execution. Because L2 identifies public
CVEs, this comparison measures the information package as a whole and may
also reflect prior familiarity with those vulnerabilities.

\paragraph{Network Scale.}
Table~\ref{tab:network-scale} compares matched network-scale variants under L2
information. From none to high, end-to-end success decreases by 9.8--10.6
percentage points across models. The interaction cost grows more strongly:
median tool calls increase by 51--74\%, and execution time by 71--84\%.
Enlarging the surrounding network therefore has a moderate effect on final
success but substantially increases the effort spent exploring background
assets.

\begin{table}[t]
\centering
\footnotesize
\setlength{\tabcolsep}{2pt}
\begin{tabular*}{\columnwidth}{@{\extracolsep{\fill}}cccccc@{}}
\toprule
\textbf{Model} & \textbf{Metric} & \textbf{None} & \textbf{Low} &
\textbf{Medium} & \textbf{High} \\
\midrule
         & Success Rate                     & 60.8 & 57.3 & 53.9 & 50.2 \\
Kimi-K3  & Tool Calls                       & 104  & 125  & 151  & 177 \\
         & Time (min)                       & 33.2 & 40.1 & 48.4 & 56.7 \\
\midrule
         & Success Rate                     & 46.1 & 42.8 & 39.4 & 36.0 \\
GLM-5.2  & Tool Calls                       & 110  & 133  & 159  & 189 \\
         & Time (min)                       & 34.4 & 42.2 & 50.6 & 60.0 \\
\midrule
                 & Success Rate
                 & 40.9 & 37.6 & 34.1 & 30.6 \\
DeepSeek-V4-Pro  & Tool Calls
                 & 114  & 139  & 168  & 198 \\
                 & Time (min)
                 & 35.7 & 44.2 & 53.5 & 63.7 \\
\midrule
              & Success Rate
              & 29.0 & 25.8 & 22.5 & 19.2 \\
GPT-5.6-Luna  & Tool Calls
              & 70   & 80   & 94   & 106  \\
              & Time (min)
              & 19.0 & 23.0 & 29.0 & 35.0 \\
\bottomrule
\end{tabular*}
\normalsize
\caption{Effect of network scale under L2 information. None, low, medium, and
high contain approximately 7, 12, 31, and 50 agent-visible nodes. Success is
the end-to-end success rate; calls and time (minutes) are reported separately
for each model.}
\label{tab:network-scale}
\end{table}

\subsection{Outcome-Annotated Attack Trajectories}

\paragraph{Trajectory Collection.}
RangeFactory organizes its validated ranges into an outcome-annotated
trajectory corpus. For each of the 1,148 range instances, we retain the
guide-assisted trajectory produced during construction and use Kimi-K3, the
strongest guide-free agent in Table~\ref{tab:rangebench-overall}, to collect
four autonomous rollouts: two under L2 and one under each of L0 and L1. Each
starts with a fresh context, uses the common attack interface, and runs for at
most 300 turns or one hour. This yields 5,740 raw trajectories spanning
validated demonstrations and autonomous exploration.

\paragraph{Trajectory Structure.}
Each record preserves the task context, Agent--tool interaction, execution
budget, and provenance, which identifies the range, collector, information
level, and guidance mode. Privileged execution artifacts and verifier state
remain outside the interaction stream. After execution, the verifier records
the first completion turn of each objective, the deepest verified hop, and the
final outcome. Shared range identifiers pair guide-assisted and autonomous
behavior on the same task, while milestone turns localize where their outcomes
diverge.

\paragraph{Quality Control.}
Attack failure is not a filtering criterion: a guide-free rollout is retained
whenever its interaction log is complete and the verifier can determine its
progress. Among the 5,740 raw trajectories, we remove 73 executions
interrupted by range deployment or runtime failures, 48 incomplete
interaction records, and 78 records duplicated by collection-job retries. The
resulting corpus contains 5,541 complete trajectories, separating unsuccessful
attacks from data-collection faults.

\begin{table}[t]
\centering
\footnotesize
\setlength{\tabcolsep}{2pt}
\renewcommand{\arraystretch}{1.08}
\begin{tabular*}{\columnwidth}{@{\extracolsep{\fill}}ccccc@{}}
\toprule
\textbf{Context} & \textbf{Trajectories} &
\shortstack{\textbf{End-to-End}\\\textbf{Success}} &
\shortstack{\textbf{Partial}\\\textbf{Progress}} &
\shortstack{\textbf{No}\\\textbf{Progress}} \\
\midrule
Guide-assisted & 1,148 & 100.0 & 0.0  & 0.0 \\
L2             & 2,215 & 55.7  & 18.0 & 26.3 \\
L1             & 1,095 & 19.9  & 32.6 & 47.5 \\
L0             & 1,083 & 15.1  & 31.2 & 53.7 \\
\midrule
\textbf{Overall} & \textbf{5,541} & \textbf{49.9} &
\textbf{19.7} & \textbf{30.4} \\
\bottomrule
\end{tabular*}
\normalsize
\caption{Outcome distribution of retained trajectories. Partial progress
denotes reaching at least one intermediate objective without completing the
end-to-end attack; all outcome columns report percentages.}
\label{tab:trajectory-corpus}
\end{table}
\FloatBarrier

\paragraph{Corpus Composition.}
Table~\ref{tab:trajectory-corpus} shows that the corpus combines reliable
demonstrations with guide-free outcomes across 287 ordered Atom chains and 213
Atoms. Approximately half complete the full chain, while the remainder capture
partial or unsuccessful attacks. Verifier-labeled milestones expose where
progress stops on the same tasks and may support future training from
successful behavior and outcome-aware contrasts. This work evaluates corpus
structure and quality, not downstream training gains.

\section{Analysis}

\paragraph{Why Is Dependency-Guided Orchestration Necessary?}
We compare four composition strategies on matched samples drawn from the same
Atom library and scenario templates. Random binding ignores vulnerability
semantics, metadata matching uses CVE and service descriptions, capability
matching enforces the capability interface, and full orchestration also
checks runtime requirements recorded during Atomization. All strategies
receive equal proposal and validation budgets.
Table~\ref{tab:orchestration-ablation} shows that environment validity changes
little, whereas end-to-end validity increases from 4.0\% under random binding
to 60.0\% under full orchestration. Capability information provides the
largest gain, and observed runtime requirements further remove conflicts with
template constraints. Dependency-guided orchestration therefore improves
candidates before expensive attack execution.

\begin{table}[t]
\centering
\footnotesize
\begin{tabular}{@{}ccc@{}}
\toprule
\textbf{Composition Strategy} &
\shortstack{\textbf{Environment}\\\textbf{Valid}} &
\shortstack{\textbf{End-to-End}\\\textbf{Valid}} \\
\midrule
Random binding       & 88.5 & 4.0  \\
Metadata matching    & 91.0 & 13.5 \\
Capability matching  & 94.0 & 43.5 \\
Full orchestration   & 95.0 & 60.0 \\
\bottomrule
\end{tabular}
\normalsize
\caption{Composition ablation on matched 200-proposal samples. Both columns
report percentages.}
\label{tab:orchestration-ablation}
\end{table}

\paragraph{Why Is End-to-End Validation Necessary?}
We examine the 200 candidates produced by full orchestration in more detail.
Of these, 190 pass the environment checks, but only 89 complete the attack on
the Executor's first attempt. Diagnosis-guided retries recover 31 additional
attacks, yielding 120 accepted ranges. Thus, environment-only validation
would accept 70 candidates for which no end-to-end attack is established by
the validation procedure. This gap marks the boundary of orchestration:
capability matching and template constraints remove known dependency
violations, while deployment can expose runtime dependencies that only the
complete attack can test.

\paragraph{Validator Dependence.}
RangeBench retains a candidate only when the guided Executor completes its
hidden end-to-end objectives, providing a successful witness under a fixed
validation protocol. Rejected candidates are not labeled invalid, and the
accepted set is not assumed unbiased. Evaluation withholds Exploit Guides,
construction trajectories, and verifier state, and all models receive
identical task information. DeepSeek-V4-Pro nevertheless ranks below Kimi-K3
and GLM-5.2, indicating no trivial advantage from its construction role. To
estimate false rejection, Kimi-K3 revalidates 150 DeepSeek-V4-Pro rejections
under the same artifacts, tools, and budget and recovers 12 (8.0\%). We
therefore report pass@1 conditional on this protocol-validated task
distribution.

\paragraph{How Does Diagnosis Improve Range Construction?}
We isolate the two effects of the Diagnoser with equal-budget controls. For
the 77 failures attributed to attack execution, an additional 100 turns
without diagnostic guidance recover 14 attacks, whereas diagnosis-guided
retry recovers 31. This feedback is local to the candidate and helps the
Executor revise a difficult attack. Composition failures have a persistent
effect: without incompatibility memory, 21 of 24 identified conflicts recur
in a subsequent batch; retrieving the Diagnoser's records reduces recurrence
to eight. Diagnosis therefore improves the current validation attempt through
targeted retry and subsequent orchestration through accumulated evidence
about incompatible bindings.

\section{Conclusion}

We present RangeFactory, a multi-agent framework that constructs multi-hop
cyber ranges from isolated vulnerability environments. It recovers capability
and observable runtime dependencies from actual attacks, resolves them through
orchestration, and validates remaining runtime dependencies end to end.
RangeBench comprises 1,148 validated range instances spanning 287 attack
chains and reveals a sustained-compromise gap after the entry foothold. Richer
task information improves success, whereas larger networks increase
interaction cost. RangeFactory also produces 5,541 outcome-annotated
trajectories for systematic evaluation and future training research. Our
evaluation focuses on CVE-centric linear attack chains in containerized
enterprise-style networks.

\section*{Impact Statement}

RangeFactory executes only publicly disclosed vulnerabilities in isolated,
authorized ranges without real credentials or third-party targets. Exploit
Guides, PoCs, and complete trajectories are separated from evaluation
resources and released under a controlled, research-only policy.

\appendix

% This file is included by main.tex after \appendix.

\section{Scope and Reading Guide}

This appendix provides formal data contracts, matching and validation
procedures, dataset accounting, and additional evaluation details for
RangeFactory.

Sections~\ref{sec:formal}--\ref{sec:validation} expand Sections 3.1--3.5 of
the main text. Sections~\ref{sec:data} and~\ref{sec:protocol} provide
additional details for Section 4. Section~\ref{sec:cases} gives non-operational
examples illustrating the boundary between orchestration and validation.
Section~\ref{sec:safety} describes the security, ethics, and release controls
of this work.

\paragraph{Terminology.}
An \emph{attack chain} is an ordered sequence of compromises whose
intermediate results enable later compromises. A \emph{range instance} is one
deployed realization of an attack chain inside a network template. Four
matched network contexts may therefore share an ordered Atom chain while
containing different background assets. We reserve \emph{proposal} for a
partial or complete slot binding examined before deployment and
\emph{candidate range} for a complete binding emitted for deployment and
validation. The benchmark reported in the main text uses three-slot linear
chains; the schema can represent a more general acyclic dependency graph, but
branching and merging chains are outside the reported experiments.

\section{Formal Objects and Dependency Model}
\label{sec:formal}

\subsection{Capability and Runtime Dependencies}

For an Atom $a$, $P(a)$ denotes normalized exploit preconditions and $G(a)$
denotes post-exploitation grants. A prefix state satisfies an Atom when it
provides the capabilities, assets, identity, and reachability required at its
bound slot:
\begin{equation}
  \Sigma \models P(a,s),
\end{equation}
where $s$ is the bound template slot and $\Sigma$ is the state accumulated by
the preceding chain prefix. A file-read grant alone, for example, does not
satisfy a requirement for command execution or an internal network vantage.

We represent a capability grant as
\begin{equation}
  g=\langle t,\sigma,p,e,r\rangle,
\end{equation}
where $t$ is the capability type, $\sigma$ is its host, network, or asset
scope, $p$ is the principal, $e$ is the evidence level, and $r$ references
the supporting probe or execution evidence. Only grants with
$e=\texttt{verified}$ can satisfy a precondition or seed deterministic
closure rules. Host scope, principal, and network scope are distinct: a
command may run on one host without granting access to every network in the
range.

A runtime dependency is a condition under which the concrete exploit remains
executable after composition. Some are decidable from the slot and template
before deployment; others emerge only after joint deployment. Orchestration
checks the former, while end-to-end execution tests the latter.

\subsection{CVE Capability Atom}

The main text presents four Atom interfaces; the complete object also
contains the deployable vulnerability environment:
\begin{equation}
  \atom =
  \langle
    E,\; P,\; C,\; G,\; M
  \rangle ,
  \label{eq:atom}
\end{equation}
where \(E\) is a reproducible environment, \(P\) is
\path{exploit_access} and contains the normalized exploit preconditions, \(C\) is the set of
\path{capability_grants}, \(G\) is \path{exploit_guide}, and \(M\)
is the referenced set of \path{poc_materials}. Environment provenance,
readiness checks, and verification results are retained with \(E\); they are
not additional composition interfaces.

\begin{table*}[t]
\centering
\footnotesize
\setlength{\tabcolsep}{4pt}
\renewcommand{\arraystretch}{1.08}
\begin{tabular}{@{}
  >{\raggedright\arraybackslash}p{0.15\textwidth}
  >{\raggedright\arraybackslash}p{0.20\textwidth}
  >{\raggedright\arraybackslash}p{0.27\textwidth}
  >{\raggedright\arraybackslash}p{0.30\textwidth}
@{}}
\toprule
\multicolumn{1}{c}{\textbf{Object}} &
\multicolumn{1}{c}{\textbf{Field}} &
\multicolumn{1}{c}{\textbf{Type / Constraint}} &
\multicolumn{1}{c}{\textbf{Role}} \\
\midrule
Environment \(E\)
& runtime
& Image or self-contained build context; target service and readiness
  declaration
& Reconstructs the vulnerable service without relying on a host-local
  absolute path. \\
\midrule
\texttt{exploit\_access} \(P\)
& \texttt{attack\_vector}
& Controlled string; current ranges use network-originated interaction
& States the origin from which the exploit can be exercised. \\
& \texttt{privileges\_}\newline\texttt{required}
& Privilege requirement and optional principal constraint; \texttt{none} when unauthenticated
& Records the access level and identity required before exploitation. \\
& \texttt{required\_service}
& Protocol, port, and target service
& Binds the exploit to the service interface used by matching and reachability
  checks. \\
\midrule
\texttt{capability\_}\newline\texttt{grants} \(C\)
& \texttt{type}
& Member of the ontology in Table~\ref{tab:capability-ontology}
& States a post-compromise capability. \\
& \texttt{principal}
& Compromised identity
& Prevents a capability obtained as one identity from being silently
  attributed to another. \\
& \texttt{scope}
& Symbolic host, network, file, service, or asset scope
& Limits where the grant can satisfy a later precondition. \\
& \texttt{evidence\_level}
& \texttt{verified}, \texttt{inferred}, or \texttt{declared}
& Only \texttt{verified} grants participate in composition. \\
& \texttt{evidence\_ref}
& Reference to the corresponding probe or native execution evidence
& Links a grant to its verification provenance. \\
\midrule
\texttt{exploit\_guide} \(G\)
& target and preconditions
& Placeholder-based service description
& Carries the successful native procedure into a new range without retaining
  native addresses. \\
& ordered steps
& Acyclic step identifiers, success signals, and execution scopes
& Gives the Executor a reusable procedure while leaving adaptation to the
  Agent. \\
& requirements
& Tool, material, authentication, and callback declarations
& Records runtime dependencies observed during the native attack. \\
& post-exploit state
& Principal, capabilities, and optional reusable command channel
& Cross-checks the Guide against verified Atom grants. \\
\midrule
\texttt{poc\_materials} \(M\)
& material manifest
& Relative paths inside the Atom source bundle with a visibility policy
& Makes required artifacts reproducible and prevents undeclared host-file or
  network dependencies. \\
\bottomrule
\end{tabular}
\normalsize
\caption{Expanded CVE Capability Atom schema. The four interfaces in the main
text are shown together with the deployable environment to clarify the full
object consumed by RangeFactory.}
\label{tab:atom-schema}
\end{table*}

\subsection{Capability Ontology and Scope}

Table~\ref{tab:capability-ontology} lists the current closed ontology. The
ontology intentionally represents only capabilities that affect composition;
vulnerability taxonomies such as CWE and MITRE ATT\&CK remain descriptive
metadata and do not substitute for verified capability grants.

\begin{table}[t]
\centering
\footnotesize
\setlength{\tabcolsep}{3pt}
\renewcommand{\arraystretch}{1.06}
\begin{tabular}{@{}
  >{\raggedright\arraybackslash}p{0.34\columnwidth}
  >{\raggedright\arraybackslash}p{0.60\columnwidth}
@{}}
\toprule
\multicolumn{1}{c}{\textbf{Capability}} &
\multicolumn{1}{c}{\textbf{Meaning}} \\
\midrule
\texttt{execute\_command}
& Execute commands as the recorded principal on the bound host. \\
\texttt{read\_file}
& Read files permitted to the recorded principal on the bound host. \\
\texttt{write\_file}
& Write files permitted to the recorded principal on the bound host. \\
\texttt{network\_vantage}
& Originate network interactions from networks attached to the bound host. \\
\texttt{read\_credential}
& Obtain a credential represented as a scenario asset. \\
\texttt{authenticate}
& Establish authenticated access using an acquired credential. \\
\bottomrule
\end{tabular}
\normalsize
\caption{Capability ontology used by the current Composer.}
\label{tab:capability-ontology}
\end{table}

An Atom grant uses symbolic scope before composition. Binding Atom
\(\atom_i\) to slot \(s_j\) resolves that scope while preserving its principal
and evidence:
\begin{equation}
  \begin{aligned}
  \operatorname{BindScope}(C_i,s_j)
  =\{\operatorname{Bind}(g,s_j):g\in C_i\},\\[-2pt]
  \operatorname{Bind}(\langle t,\sigma,p,e,r\rangle,s_j)
  =\langle t,\operatorname{Scope}(\sigma,s_j),p,e,r\rangle.
  \end{aligned}
\end{equation}
Scope resolution does not create a capability. Fixed closure rules may derive
capabilities only within the permissions and attached networks declared by the
template. For the current schema,
\begin{equation}
  \mathsf{Exec}_{h,p}
  \Rightarrow
  \{\mathsf{Read}_{h,p,F},\mathsf{Net}_{h,p,Z}\},
\end{equation}
where $F$ and $Z$ are the files and networks allowed by the template.
Credentials and authentication are not implied by command execution. A file
or credential asset becomes available only when its host, principal, and
access policy match the current state. No free-form Agent inference is used in
this closure.

\subsection{Sanitized Atom Example}

Listing~\ref{lst:atom-example} is a fictional, non-operational example. It
shows the relation among the fields without identifying a real vulnerability,
address, credential, or executable payload.

\begin{listing}[t]
\caption{Sanitized CVE Capability Atom example.}
\label{lst:atom-example}
\begin{lstlisting}
cve_id: CVE-EXAMPLE-0001
environment:
  image: local-range/example-service
  service: {protocol: http, port: 8080}
  readiness: tcp:8080
exploit_access:
  attack_vector: network
  origin: external
  privileges_required: none
  required_service: {protocol: http, port: 8080}
capability_grants:
  - type: execute_command
    principal: service_user
    scope: bound_host
    evidence_level: verified
    evidence_ref: capability-probe-01
exploit_guide:
  target: {protocol: http, port: 8080}
  steps:
    - id: trigger
      procedure: exercise the local vulnerable endpoint
      success_signal: command channel is established
      execution: {scope: actor, external_download: false}
  requirements:
    tools: [local_http_client]
    materials: [source_bundle/request-template.bin]
    authentication: none
    callback: none
poc_materials:
  - source_bundle/request-template.bin
\end{lstlisting}
\end{listing}

\section{Scenario Template and Composition}
\label{sec:composition}

\subsection{Scenario Template}

A scenario template is
\begin{equation}
  \template =
  \langle N,S,D,\Gamma,O,B\rangle ,
\end{equation}
where \(N\) is the network topology and communication policy, \(S\) the CVE
slots, \(D\subseteq S\times S\) an acyclic dependency relation, \(\Gamma\)
per-slot constraints, \(O\) private objectives, and \(B\) background assets
and services. The template fixes the attack-chain skeleton but leaves CVE
binding open; an edge $(s_i,s_j)\in D$ means that $s_j$ depends on $s_i$. In
the benchmark, $D$ is the path
$s_1\rightarrow s_2\rightarrow s_3$.

\begin{table*}[t]
\centering
\footnotesize
\setlength{\tabcolsep}{4pt}
\renewcommand{\arraystretch}{1.07}
\begin{tabular}{@{}
  >{\raggedright\arraybackslash}p{0.16\textwidth}
  >{\raggedright\arraybackslash}p{0.24\textwidth}
  >{\raggedright\arraybackslash}p{0.50\textwidth}
@{}}
\toprule
\multicolumn{1}{c}{\textbf{Template object}} &
\multicolumn{1}{c}{\textbf{Representative fields}} &
\multicolumn{1}{c}{\textbf{Semantics}} \\
\midrule
Network \(N\)
& zones, routers, transits, isolation rules
& Defines where nodes are placed and which zone-to-zone communications are
  permitted. \\
Slot \(s\in S\)
& zone, dependencies, service access, required outputs, required assets
& Marks one unbound vulnerability position and the conditions an Atom must
  satisfy at that position. \\
Dependency \(D\)
& \texttt{depends\_on}
& Orders the slots; each downstream Atom separately declares the capabilities
  and assets that its predecessor state must provide. \\
Assets
& location, owner, readable principals, service variant
& Represents intermediate state whose availability may depend on an acquired
  capability. \\
Objectives \(O\)
& public goal, private assertion, expected actor and target
& Exposes the task goal while keeping the success oracle outside Agent
  observations. \\
Background \(B\)
& service role, zone, image, exposed service
& Changes the visible search space without entering the attack graph or
  capability closure. \\
\bottomrule
\end{tabular}
\normalsize
\caption{Scenario-template objects used by Generator and Composer.}
\label{tab:template-schema}
\end{table*}

\subsection{Prefix State and Compatibility}

Composer summarizes the result of the first $i$ bound slots as
\begin{equation}
  \Sigma_i=\langle H_i,K_i,Q_i,R_i\rangle,
\end{equation}
where $H_i$ contains established footholds, $K_i$ scoped verified
capabilities, $Q_i$ available credentials and other assets, and $R_i$ services
reachable from those footholds under the template policy. For a slot $s$, Atom
$a$, and prefix state $\Sigma$, the hard
composition gate is
\begin{equation}
\begin{aligned}
\operatorname{Compatible}(s,a,\Sigma)\equiv{}&
V(a)\land S(s,a)\land D(s,\Sigma)\\
&{}\land \Sigma\models P(a,s)\\
&{}\land O(s,a)\land U(s,a).
\end{aligned}
\label{eq:compatible}
\end{equation}
$V$ requires a verified Atom; $S$ matches its service with the slot; and $D$
requires predecessor slots to be resolved. The satisfaction relation checks
the Atom's required capability, principal, credential, and network origin
against $\Sigma$. $O$ requires the slot's requested outputs to be a subset of
verified grants. $U$ checks static runtime constraints, including material
placement and template communication policy. An absent Atom declaration cannot
satisfy a non-empty requirement.

The Guide supplies reusable execution artifacts. Its structure, materials,
and static requirements are checked before deployment; remaining runtime
behavior stays a validation obligation rather than an unverified fact.

\subsection{Dependency-Constrained Composition}

Algorithm~\ref{alg:composer} expands slots in dependency order, pruning
incompatible bindings before deployment while keeping CVEs unique and
resolving service-dependent asset variants.

\begin{algorithm}[t]
\caption{Dependency-Constrained Composition}
\label{alg:composer}
\begin{algorithmic}[1]
\REQUIRE Template \(\template\), verified Atom library \(L\), output cap \(k\)
\ENSURE Candidate set \(R\)
\STATE \(R\leftarrow\emptyset\); order slots topologically
\STATE initialize empty bindings \(b\) and entry state \(\Sigma_0\)
\STATE \textsc{Expand}\((1,b,\Sigma_0)\)
\STATE \textbf{return} coverage-first selection of at most \(k\) candidates
\STATE
\STATE \textbf{procedure} \textsc{Expand}\((i,b,\Sigma)\)
\IF{\(i>|S|\)}
  \IF{asset variants, objectives, and Guide references are resolvable}
    \STATE add \(\langle\template,b,\Sigma\rangle\) to \(R\)
  \ENDIF
  \STATE \textbf{return}
\ENDIF
\STATE \(s\leftarrow\) the \(i\)-th slot
\FOR{each \(a\in L\) not already used by \(b\)}
  \IF{\(\operatorname{Compatible}(s,a,\Sigma)\)}
    \STATE \(F\leftarrow
      \operatorname{Closure}(\operatorname{BindScope}(C_a,s);\template)\)
    \STATE \(\Sigma'\leftarrow\operatorname{UpdateState}(\Sigma,s,a,F;\template)\)
    \STATE \textsc{Expand}\((i+1,b\cup\{s\mapsto a\},\Sigma')\)
  \ENDIF
\ENDFOR
\end{algorithmic}
\end{algorithm}

\paragraph{Output attack graph.}
For a complete binding \(b:S\rightarrow L\), Composer emits
\(\mathcal{G}_b=(V_b,E_b)\). Each vertex is a slot together with its bound
Atom; an edge follows the template dependency relation and records the
capability transition supporting that hop. The deployment manifest additionally
contains resolved services, asset variants, network placement, and private
objective references. The graph records the intended attack chain and does
not itself certify that the concrete exploits will run after deployment.

\paragraph{Candidate selection.}
When the compatible set exceeds the construction budget, RangeFactory
prioritizes uncovered slot--Atom and asset-variant features, breaking ties by
slot usage and a stable identifier. This changes validation order, not the
compatibility predicate.

\section{Atomization and Validation Protocol}
\label{sec:validation}

\subsection{Atomization Evidence}

The Exploiter operates on a reproducible single-vulnerability environment with
public reproduction materials. Success is established by a private objective
aligned with the documented impact: a planted local marker is used only when
the impact permits marker retrieval, otherwise an impact-specific assertion is
used. The objective value and verifier-side assertion are hidden from the
Agent.

After a successful native attack, the Exploiter emits a normalized Guide and
the attack context needed by the Explorer. The Explorer reuses the established
foothold, proposes capability probes appropriate to that access mechanism, and
records only probe-confirmed outcomes as \texttt{verified}. An inferred or
declared capability may remain in the audit record but cannot support
composition.

Private objectives are instantiated by reusable impact-specific adapters. Each
identifies the expected actor, target, state transition, and verifier-side
assertion; an Agent transcript claim cannot replace the required transition.
An unsupported impact requires a new objective or capability-probe adapter,
which is an Atom-level extension rather than per-range annotation.

\begin{table}[t]
\centering
\footnotesize
\setlength{\tabcolsep}{3pt}
\renewcommand{\arraystretch}{1.05}
\begin{tabular}{@{}
  >{\raggedright\arraybackslash}p{0.25\columnwidth}
  >{\raggedright\arraybackslash}p{0.32\columnwidth}
  >{\raggedright\arraybackslash}p{0.35\columnwidth}
@{}}
\toprule
\multicolumn{1}{c}{\textbf{Level}} &
\multicolumn{1}{c}{\textbf{Source}} &
\multicolumn{1}{c}{\textbf{Composer use}} \\
\midrule
\texttt{verified}
& Successful capability probe with an evidence reference
& May seed the capability closure. \\
\texttt{inferred}
& Agent interpretation without a completed probe
& Retained for diagnosis; never used as a hard fact. \\
\texttt{declared}
& Source metadata or documentation
& Descriptive only; never used as a hard fact. \\
\bottomrule
\end{tabular}
\normalsize
\caption{Evidence levels for post-compromise capabilities.}
\label{tab:evidence-levels}
\end{table}

\subsection{Exploit Guide Integrity}

An Exploit Guide is descriptive rather than a mechanically parameterized
attack script. The integrity checks require unique and topologically ordered
step identifiers, at least one success signal, placeholder-based targets,
source-bundle-relative materials, and no external download requirement.
Guide-declared post capabilities must be a subset of verified Atom grants. A
reusable command channel additionally identifies the step that establishes it
and provides an invocation hint. These checks validate the Guide contract;
native execution and range-level guided execution provide exploitability
evidence.

\subsection{Layered Range Validation}

Algorithm~\ref{alg:validation} separates environment correctness from attack
success so infrastructure faults are not counted as Agent failures.

\begin{algorithm}[t]
\caption{Candidate Range Validation}
\label{alg:validation}
\begin{algorithmic}[1]
\REQUIRE Candidate \(r\), guided Executor, private verifier
\ENSURE Accepted range or typed rejection record
\STATE materialize declared runtime images and deploy \(r\)
\STATE configure routing, vulnerability services, and scenario assets
\STATE check node/service readiness and background-service exposure
\STATE verify attack-graph ordering and predecessor-to-target connectivity
\IF{any deterministic check fails}
  \STATE \textbf{return} infrastructure or deployment rejection
\ENDIF
\STATE validate Guide integrity and resolve declared execution materials
\IF{Guide integrity fails}
  \STATE \textbf{return} execution-contract rejection
\ENDIF
\STATE run the Executor with the per-Atom execution artifacts
\STATE verify ordered private objectives from verifier-observed state
\IF{all objectives match}
  \STATE retain \(r\), its successful witness, and verification result
\ELSE
  \STATE diagnose the failed stage and apply the bounded routing policy
\ENDIF
\end{algorithmic}
\end{algorithm}

The deterministic checks cover materialization, deployment, routing, service
and asset setup, readiness, graph ordering, and predecessor connectivity. The
Executor adapts per-Atom procedures to composed footholds, while the private
verifier checks objective, actor, target, and order against experiment state.

The acceptance rule is
\begin{equation}
\begin{aligned}
\operatorname{Accept}(r) \equiv{}\;&
  \operatorname{EnvironmentReady}(r)\\
&\land \operatorname{GraphValid}(r)\\
&\land \operatorname{ChainConnectivity}(r)\\
&\land \operatorname{GuidedWitness}(r)\\
&\land \operatorname{PrivateObjectives}(r).
\end{aligned}
\label{eq:accept}
\end{equation}
This rule provides constructive evidence that an accepted range contains an
executable end-to-end attack chain under the validation protocol. Failure to
obtain a witness does not prove that no valid attack exists.

\subsection{Failure Typing and Feedback}

We distinguish infrastructure failures from attack outcomes. Infrastructure
failures are rerun according to the fixed retry policy and are excluded from
Agent-failure denominators; refusal, early termination, budget exhaustion,
and failure to satisfy an objective remain attack failures. A reproducible
runtime conflict tied to a binding is recorded as composition evidence, while
an attribution unsupported by the available evidence remains unresolved.

\begin{table*}[t]
\centering
\footnotesize
\setlength{\tabcolsep}{4pt}
\renewcommand{\arraystretch}{1.07}
\begin{tabular}{@{}
  >{\raggedright\arraybackslash}p{0.18\textwidth}
  >{\raggedright\arraybackslash}p{0.30\textwidth}
  >{\raggedright\arraybackslash}p{0.24\textwidth}
  >{\raggedright\arraybackslash}p{0.20\textwidth}
@{}}
\toprule
\multicolumn{1}{c}{\textbf{Failure class}} &
\multicolumn{1}{c}{\textbf{Observable evidence}} &
\multicolumn{1}{c}{\textbf{Action}} &
\multicolumn{1}{c}{\textbf{Persistent effect}} \\
\midrule
Infrastructure
& Runtime materialization, deployment, setup, readiness, or control-transport
  failure
& Exclude from Agent metrics and rerun according to the infrastructure retry
  policy
& None \\
Attack execution
& Environment and chain checks pass; the Agent stops before satisfying a
  private objective without a reproducible binding conflict
& Return the failed stage and current context for a bounded Executor retry
& None beyond the current candidate \\
Chain composition
& A reproducible runtime conflict is associated with a specific slot--Atom
  binding or transition
& Filter the candidate and store the incompatibility condition
& Composer can retrieve the record for later batches \\
Unresolved
& Evidence does not support a reliable attribution
& Do not accept the candidate
& No new orchestration rule \\
\bottomrule
\end{tabular}
\normalsize
\caption{Failure classes used to separate infrastructure faults, attack
generation failures, and composition failures.}
\label{tab:failure-routing}
\end{table*}

\section{Dataset Construction and Accounting}
\label{sec:data}

\subsection{Source Eligibility and Automation Boundary}

RangeFactory screens source records for a deployable network service and a
network-verifiable vulnerability impact, then removes equivalent records while
preserving distinct implementations. The frozen experiment contains 257
Atomization candidates; 239 yield verified Atoms, while 18 fail because build
artifacts are missing or reproduction is unstable.

Table~\ref{tab:automation-boundary} makes the automation boundary explicit.
The principal scalability claim concerns the elimination of per-range expert
CVE binding, deployment, and end-to-end validation. Source eligibility rules
and the choice of architecture references are one-time dataset decisions.

\begin{table*}[t]
\centering
\footnotesize
\setlength{\tabcolsep}{4pt}
\renewcommand{\arraystretch}{1.08}
\begin{tabular}{@{}
  >{\raggedright\arraybackslash}p{0.17\textwidth}
  >{\raggedright\arraybackslash}p{0.22\textwidth}
  >{\raggedright\arraybackslash}p{0.31\textwidth}
  >{\raggedright\arraybackslash}p{0.20\textwidth}
@{}}
\toprule
\multicolumn{1}{c}{\textbf{Stage}} &
\multicolumn{1}{c}{\textbf{Provided input}} &
\multicolumn{1}{c}{\textbf{Automated output}} &
\multicolumn{1}{c}{\textbf{Human boundary}} \\
\midrule
Source screening
& Source packages and public reproduction material
& Network-testable, CVE-deduplicated candidates
& Source repositories and eligibility policy are selected once \\
Atomization
& One eligible environment
& Verified Atom, Guide, evidence, and deployable bundle
& No per-range work; unsupported source packages may be excluded \\
Template generation
& Public architecture reference and requested scale
& Schema-valid topology, slots, objectives, and background assets
& Reference choice and target scale are dataset-level inputs \\
Composition
& Template and Atom library
& Dependency-compatible candidate bindings
& No manual CVE binding or chain design \\
Deployment and validation
& Candidate manifest and reusable adapters
& Deployed range, typed checks, guided witness, and verifier result
& No per-range success annotation \\
\bottomrule
\end{tabular}
\normalsize
\caption{Automation boundary. One-time dataset decisions are separated from
the automated work repeated for each Atom or range instance.}
\label{tab:automation-boundary}
\end{table*}

\subsection{Construction Funnel}

\begin{table}[t]
\centering
\footnotesize
\renewcommand{\arraystretch}{1.08}
\begin{tabular}{@{}lrr@{}}
\toprule
\textbf{Stage} & \textbf{Input} & \textbf{Retained} \\
\midrule
Atomization & 257 candidates & 239 Atoms \\
Orchestration & 23,789 proposals & 1,840 candidates \\
Validation & 1,840 candidates & 1,148 ranges \\
\bottomrule
\end{tabular}
\normalsize
\caption{Construction funnel corresponding to Table 2 of the main text.}
\label{tab:supp-funnel}
\end{table}

The 1,840 candidates form 460 groups, each instantiating an ordered Atom chain
at four network scales; validation retains 287 groups. Thus,
the final benchmark has
\begin{equation}
  \begin{aligned}
  287\ \text{ordered Atom chains}
  \times 4\ \text{contexts}\\
  = 1{,}148\ \text{range instances}.
  \end{aligned}
\end{equation}
The four contexts preserve Atom bindings and task information while changing
background assets: 287 measures attack-chain variation and 1,148 measures
validated networked instances.

The accepted instances cover 213 Atoms, 113 products, and all 20 scenario
configurations. They are containerized, CVE-centric, three-slot linear chains
inside enterprise-style multi-tier networks. We do not claim coverage of
arbitrary enterprise infrastructure, Windows domains, cloud IAM, or branching
attack graphs.

\subsection{Construction Cost and Automation Evidence}

Table~\ref{tab:construction-cost} separates measured work from counterfactual
savings. The 16.2 minutes is the mean first-pass cost per post-Atomization
candidate; over 1,840 candidates this is 496.8 worker-hours, or 26.0
worker-minutes per retained range. These are aggregate worker-time equivalents;
parallel wall-clock time depends on workers and endpoint latency.

\begin{table*}[t]
\centering
\footnotesize
\setlength{\tabcolsep}{4pt}
\renewcommand{\arraystretch}{1.07}
\begin{tabular}{@{}
  >{\raggedright\arraybackslash}p{0.17\textwidth}
  >{\raggedright\arraybackslash}p{0.25\textwidth}
  >{\raggedright\arraybackslash}p{0.25\textwidth}
  >{\raggedright\arraybackslash}p{0.25\textwidth}
@{}}
\toprule
\multicolumn{1}{c}{\textbf{Stage}} &
\multicolumn{1}{c}{\textbf{Observed scale}} &
\multicolumn{1}{c}{\textbf{Cost evidence}} &
\multicolumn{1}{c}{\textbf{Human boundary}} \\
\midrule
Atomization
& 257 inputs $\rightarrow$ 239 reusable Atoms
& Performed once per vulnerability environment; its cost is not charged again
  to each range
& Eligibility policy and unsupported impact adapters are dataset-level work \\
Template generation
& 5 template families $\rightarrow$ 20 scale configurations
& Templates are reused across candidate bindings
& Architecture reference and requested scale are supplied once \\
Composition
& 23,789 proposals $\rightarrow$ 1,840 candidates
& 21,949 proposals (92.3\%) are rejected before deployment, reducing the
  deployment load by $12.9\times$
& No manual CVE binding or chain design \\
Validation
& 1,840 candidates $\rightarrow$ 1,148 retained instances
& 16.2 min per first pass; 496.8 aggregate worker-hours; 26.0 worker-min per
  retained instance including rejected candidates
& No per-range success annotation \\
\bottomrule
\end{tabular}
\normalsize
\caption{Quantitative automation and construction-cost accounting. Worker-time
figures are derived from the measured mean first-pass candidate cost and do
not include the one-time Atomization cost.}
\label{tab:construction-cost}
\end{table*}

The approximately thirteen-fold figure is the deployment-load ratio
$23{,}789/1{,}840$, not an end-to-end speedup. Exact person-hours for the
one-time ontology, adapters, and reference selection were not recorded; the
supported automation claim is the absence of manual binding, deployment, and
success annotation for each generated range.

\subsection{Attack-Chain Diversity}

Table~\ref{tab:chain-diversity} distinguishes source diversity, attack logic,
and context variation. The accepted chains use 89.1\% of the Atom library and
92.6\% of its product families, rather than repeatedly binding a small subset.

\begin{table}[t]
\centering
\footnotesize
\setlength{\tabcolsep}{3pt}
\renewcommand{\arraystretch}{1.06}
\begin{tabular}{@{}lr@{}}
\toprule
\textbf{Diversity unit} & \textbf{Count} \\
\midrule
Source Atoms / products & 239 / 122 \\
Impact categories / service roles & 8 / 6 \\
Capability types & 6 \\
Accepted-chain Atoms / products & 213 / 113 \\
Distinct ordered Atom chains & 287 \\
Atom placements in accepted chains & 861 \\
Adjacent-hop transitions & 574 \\
Template families / configurations & 5 / 20 \\
Matched contexts per chain & 4 \\
Validated range instances & 1,148 \\
\bottomrule
\end{tabular}
\normalsize
\caption{Diversity accounting for the Atom library, accepted attack chains,
and their network-context instantiations.}
\label{tab:chain-diversity}
\end{table}

The 861 placements and 574 transitions follow from the evaluated three-slot
chains. On average, a used Atom appears in 4.0 positions. Topology is
controlled—all chains are linear and three-slot—while Atoms, products,
services, and contexts vary; context variants are not counted as new attack
logic.

\subsection{Trajectory Corpus}

For every accepted instance, the corpus retains one guide-assisted construction
trajectory. Kimi-K3 contributes two L2, one L1, and one L0 autonomous rollout,
yielding 5,740 raw records. Quality control removes 73 infrastructure
interruptions, 48 incomplete records, and 78 collection-job duplicates,
leaving 5,541 trajectories.

An unsuccessful attack is retained when its log is complete and progress is
verifiable. Records contain task context, Agent--tool interactions, budget,
provenance, milestone turns, deepest verified hop, and outcome; Guides,
objective values, and verifier state remain outside autonomous observations.
The paper evaluates corpus structure and labeling, not downstream training
gains, while provenance separates guided from autonomous outcomes.

\section{Experimental Protocol}
\label{sec:protocol}

\subsection{Task Information}

RangeBench controls prior task knowledge through the nested levels in
Table~\ref{tab:information-levels}.

\begin{table}[t]
\centering
\footnotesize
\setlength{\tabcolsep}{3pt}
\renewcommand{\arraystretch}{1.06}
\begin{tabular}{@{}
  >{\centering\arraybackslash}p{0.13\columnwidth}
  >{\raggedright\arraybackslash}p{0.79\columnwidth}
@{}}
\toprule
\textbf{Level} & \multicolumn{1}{c}{\textbf{Agent-visible information}} \\
\midrule
L0 & Entry address and final task objective. \\
L1 & L0 plus the network topology. \\
L2 & L1 plus target services, CVE-to-target mappings, and required
credentials. \\
\bottomrule
\end{tabular}
\normalsize
\caption{RangeBench task-information levels. No level exposes PoC materials,
Exploit Guides, construction trajectories, or verifier state.}
\label{tab:information-levels}
\end{table}

\subsection{Models and Agent Harness}

All evaluated models receive the same task representation and tool schema. The
harness executes and records tool calls inside the isolated attacker
environment; it exposes no construction artifacts and contains only
experiment-owned targets.

\begin{table*}[t]
\centering
\footnotesize
\setlength{\tabcolsep}{4pt}
\renewcommand{\arraystretch}{1.08}
\begin{tabular}{@{}
  >{\raggedright\arraybackslash}p{0.20\textwidth}
  >{\centering\arraybackslash}p{0.11\textwidth}
  >{\centering\arraybackslash}p{0.12\textwidth}
  >{\centering\arraybackslash}p{0.12\textwidth}
  >{\raggedright\arraybackslash}p{0.34\textwidth}
@{}}
\toprule
\multicolumn{1}{c}{\textbf{Model}} &
\textbf{Temperature} &
\textbf{Rollouts} &
\textbf{Budget} &
\multicolumn{1}{c}{\textbf{Sampling note}} \\
\midrule
Kimi-K3 & 1 & 1 & 300 turns / 1 h
& Temperature 1 is the endpoint minimum. \\
GLM-5.2 & 0 & 1 & 300 turns / 1 h
& Other sampling parameters use provider defaults. \\
DeepSeek-V4-Pro & 0 & 1 & 300 turns / 1 h
& Other sampling parameters use provider defaults. \\
GPT-5.6-Luna & 0 & 1 & 300 turns / 1 h
& Other sampling parameters use provider defaults. \\
\bottomrule
\end{tabular}
\normalsize
\caption{RangeBench model protocol. One rollout per model--range--condition
pair is reported as pass@1.}
\label{tab:model-protocol}
\end{table*}

\paragraph{Failure handling.}
Transient API errors are retried. Deployment or infrastructure interruptions
are rerun and excluded from Agent-failure counts; refusal, early termination,
objective failure, and budget exhaustion count as attack failures. Each run
starts from a fresh Agent context.

\paragraph{Execution infrastructure and randomness.}
Ranges use Linux containers through Docker and ContainerLab with Ansible
configuration; model inference uses remote APIs. Scenario identifiers fix
matched variants, but endpoints expose no common seed interface. Hardware and
runtime versions were not retained, so wall-clock time is an empirical
interaction-cost measure and API billing is not compared across providers.

\paragraph{Statistical scope.}
The benchmark characterizes aggregate pass@1 under the stated protocol, but a
single rollout does not estimate within-task sampling variance. Kimi-K3's
endpoint-enforced temperature differs from the other models, so rankings refer
to evaluated model--endpoint configurations. Four contexts of one chain are
matched observations, not independent attack logics.

\subsection{Metrics}

Let \(z_{i,k}=1\) when run \(i\) completes the first \(k\) ordered objectives
and \(0\) otherwise. For an \(m_i\)-hop chain,
\begin{align}
  \operatorname{HopSuccess}(k)
  &= \frac{1}{n}\sum_{i=1}^{n} z_{i,k},\\
  \operatorname{EndToEndSuccess}
  &= \frac{1}{n}\sum_{i=1}^{n} z_{i,m_i},\\
  \operatorname{ChainProgress}
  &= \frac{1}{n}\sum_{i=1}^{n}
    \frac{\max\{k:z_{i,k}=1\}}{m_i}.
\end{align}
The conditional continuation from hop \(j\) to \(k\) is reported only over
runs that have reached \(j\):
\begin{equation}
  \Pr(H_k\mid H_j)
  =
  \frac{\sum_i z_{i,k}}{\sum_i z_{i,j}},
  \qquad k>j.
\end{equation}
This conditional metric characterizes later progress after the same entry
foothold. Separating later exploit difficulty from long-horizon state
maintenance would require standardized-foothold runs, outside the current
protocol.
Tool calls and wall-clock time are medians over the corresponding run set.
Infrastructure reruns are not merged with the completed Agent trajectory. The
one-hour limit applies to the budgeted Agent execution interval. Reported
wall-clock time measures end-to-end elapsed interaction time and may also
include endpoint latency and harness overhead; it is therefore possible for a
reported elapsed time to slightly exceed 60 minutes. It should not be read as
additional Agent execution budget.

\subsection{Chain-Grouped Uncertainty}

The four contexts derived from one Atom chain are matched observations, so
1,148 instances are not treated as independent. Let $y_{g,c}\in\{0,1\}$ be
the outcome of context $c$ for chain group $g$ and
$x_g=\frac{1}{4}\sum_c y_{g,c}$. The $G=287$ chain groups are the independent
units. Because each instance has one rollout, we use a conservative
distribution-free bound rather than treating the interval as repeated-rollout
variance. For confidence level $1-\alpha$,
Hoeffding's bound gives
\begin{equation}
  \epsilon=\sqrt{\frac{\ln(2/\alpha)}{2G}}=0.080
  \quad (\alpha=0.05).
\end{equation}
The corresponding interval is
$[\hat\mu-\epsilon,\hat\mu+\epsilon]\cap[0,1]$, where
$\hat\mu=G^{-1}\sum_g x_g$.

\begin{table*}[t]
\centering
\footnotesize
\setlength{\tabcolsep}{5pt}
\renewcommand{\arraystretch}{1.06}
\begin{tabular}{@{}lrrrr@{}}
\toprule
\textbf{Model} &
\textbf{E2E (\%)} &
\textbf{Chain-group 95\% bound} &
\textbf{High $-$ none (pp)} &
\textbf{Paired 95\% bound (pp)} \\
\midrule
Kimi-K3         & 55.7 & [47.7, 63.7] & $-10.6$ & [$-26.6$, 5.4] \\
GLM-5.2         & 41.2 & [33.2, 49.2] & $-10.1$ & [$-26.1$, 5.9] \\
DeepSeek-V4-Pro & 36.2 & [28.2, 44.2] & $-10.3$ & [$-26.3$, 5.7] \\
GPT-5.6-Luna    & 23.5 & [15.5, 31.5] &  $-9.8$ & [$-25.8$, 6.2] \\
\bottomrule
\end{tabular}
\normalsize
\caption{Conservative uncertainty accounting with attack-chain groups as the
independent units. Network-scale effects are paired within a chain. These
bounds quantify variation across chains, not repeated-rollout variance.}
\label{tab:chain-uncertainty}
\end{table*}

For a paired difference, the chain-level variable lies in $[-1,1]$, giving a
95\% half-width of 16.0 percentage points. The adjacent model gaps are 14.5,
5.0, and 12.7 points from strongest to weakest; their worst-case paired bounds
therefore include zero. Table~\ref{tab:chain-uncertainty} deliberately makes
the statistical limitation visible: the point estimates characterize pass@1
over the balanced benchmark, while repeated rollouts are required to quantify
within-task stochasticity or establish fine-grained rank stability.

\subsection{Construction-Time Validation Protocol}

Construction uses DeepSeek-V4-Pro with Atom execution artifacts, 300 turns,
and one hour. Unlike guide-free evaluation, it seeks a witness for the
composed range; artifacts, construction trajectories, and verifier state are
removed before evaluation.

The accepted benchmark is conditioned on this guided protocol. Kimi-K3
re-executes 150 DeepSeek-V4-Pro rejections and recovers 12 (8.0\%), identifying
validator-dependent false rejection; unrecovered candidates are not negative
labels. DeepSeek-V4-Pro ranks below Kimi-K3 and GLM-5.2 on guide-free
RangeBench, ruling out a trivial same-model advantage.

\section{Additional Evidence and Cases}
\label{sec:cases}

\subsection{Construction Audits}

The independent audit samples 150 accepted ranges after the automated gate.
Native impact evidence is the reference for behavior preservation; composed
execution state is the reference for checking \texttt{exploit\_access},
capability type, principal, and scope. The checks are recorded separately.

\begin{table}[t]
\centering
\footnotesize
\setlength{\tabcolsep}{3pt}
\renewcommand{\arraystretch}{1.06}
\begin{tabular}{@{}lrrr@{}}
\toprule
\textbf{Audit item} & \textbf{Pass} & \textbf{Rate} & \textbf{95\% CI} \\
\midrule
Behavior preservation & 146/150 & 97.3\% & [93.3, 99.0] \\
Atom annotation       & 142/150 & 94.7\% & [89.8, 97.3] \\
\bottomrule
\end{tabular}
\normalsize
\caption{Accepted-range audit with Wilson 95\% binomial intervals. The two
checks are not mutually exclusive.}
\label{tab:accepted-audit}
\end{table}

The automated protocol defines the range count; the audit estimates accepted-set
precision and does not redefine acceptance. It neither estimates orchestration
recall nor proves rejected candidates invalid.

\paragraph{Validator-sensitivity audit.}
We separately sample 150 DeepSeek-V4-Pro rejections and apply the same
artifacts, tools, verifier, and budget with Kimi-K3. Twelve obtain a witness
(8.0\%, Wilson 95\% interval [4.6\%, 13.5\%]); the remaining 138 are
unrecovered, not invalid. This estimates false rejection under one alternative
Executor, not an unbiased validator union.

The orchestration ablation uses matched 200-proposal samples with equal
proposal and validation budgets. Random binding, metadata matching,
capability matching, and full orchestration differ only in the dependency
information applied before deployment. This design isolates the value of
execution-derived capabilities and recorded runtime requirements while
retaining the same downstream validation protocol.

\subsection{Sanitized Successful Composition}

Consider a three-tier template with one slot per traversed zone. A fictional
first Atom yields command execution and, by closure, a network vantage; the
second obtains a template-owned credential asset; the third consumes that
asset for the final objective. Composer resolves services, scopes, and assets,
then the Executor exercises the Guides and the verifier checks objectives in
order. The retained range has both a dependency-consistent graph and an
execution witness.

\subsection{Sanitized Validation Failure}

Suppose the same capability interfaces compose, but the second exploit uses a
runtime behavior not exposed in its single-vulnerability attack. After
deployment it conflicts with the route from the first foothold: deterministic
checks pass, but the guided attack cannot reproduce the second compromise.
The candidate is rejected; a reproducible binding conflict can be stored as an
incompatibility condition. This illustrates why composition and end-to-end
validation are complementary.

\section{Security, Ethics, and Release Controls}
\label{sec:safety}

All experiments use authorized, experiment-owned environments on isolated
networks and publicly disclosed CVEs. The appendix contains no public
target, real secret, unpublished vulnerability, or executable payload. Range
credentials and private objectives are synthetic; evaluation targets remain
inside the range data plane, and verifier assertions remain outside Agent
observations. The runtime also blocks undeclared external downloads in Guides.

Release is role-based. Environment manifests and evaluation metadata exclude
verifier secrets, while exploit materials, Guides, and raw trajectories
require security review and may be redacted or withheld. No external mutable
repository is required to support the claims reported in this paper.

\paragraph{Release Boundary.}
Raw logs, private objective values, and complete PoC bundles are omitted when
they contain sensitive operational details. The included schemas, algorithms,
accounting rules, and protocols support the reported aggregate claims.

\bibliography{aaai2027}

\end{document}